\documentclass[final,3p,sort&compress]{elsarticle}

\usepackage{graphicx}

\usepackage{amsmath,amssymb,amsfonts}
\usepackage{bm}

\def\be{\begin{equation}}
\def\ee{\end{equation}}
\def\ba{\begin{eqnarray}}
\def\ea{\end{eqnarray}}
\def\bd{\begin{displaymath}}
\def\ed{\end{displaymath}}
\def\bq{\begin{eqnarray}}
\def\eq{\end{eqnarray}}

\journal{Annals of Physics}

\begin{document}

\begin{frontmatter}

\title{Quantum Zeno effect: Quantum shuffling and Markovianity}

\author{A. S. Sanz\corref{corresp}} \ead{asanz@iff.csic.es}

\author{C. Sanz-Sanz}

\author{T. Gonz\'alez-Lezana}

\author{O. Roncero}

\author{S. Miret-Art\'es}

\cortext[corresp]{Corresponding author}

\address{Instituto de F\'{\i}sica Fundamental (IFF--CSIC),
Serrano 123, 28006 Madrid, Spain}

\begin{abstract}
The behavior displayed by a quantum system when it is perturbed by a
series of von Neumann measurements along time is analyzed.
Because of the similarity between this general process with giving a
deck of playing cards a shuffle, here it is referred to as {\it quantum
shuffling}, showing that the quantum Zeno and anti-Zeno effects emerge
naturally as two time limits.
Within this framework, a connection between the gradual transition from
anti-Zeno to Zeno behavior and the appearance of an underlying
Markovian dynamics is found.
Accordingly, although {\it a priori} it might result counterintuitive,
the quantum Zeno effect corresponds to a dynamical regime where any
trace of knowledge on how the unperturbed system should evolve
initially is wiped out (very rapid shuffling).
This would explain why the system apparently does not evolve or
decay for a relatively long time, although it eventually undergoes
an exponential decay.
By means of a simple working model, conditions characterizing the
shuffling dynamics have been determined, which can be of help to
understand and to devise quantum control mechanisms in a number of
processes from the atomic, molecular and optical physics.
\end{abstract}



\end{frontmatter}



\section{Introduction}
\label{sec1}

The legacy of Zeno of Elea becomes very apparent through calculus, the
pillar of physics.
It is not difficult to find manifestations of his famous
paradoxes throughout the subtleties of any of our physical theories
\cite{mazur-bk:2007}.
In quantum mechanics, for example, Zeno's paradox of the arrow is of
particular interest, for it has given rise to what is now known as
quantum Zeno effect (QZE), which constitutes an active field of
research \cite{pascazio:JPA:2008}.
As conjectured by Misra and Sudarshan \cite{misra:JMathPhys:1977}, this
effect essentially consists of inhibiting the evolution of an unstable
quantum system by a succession of shortly-spaced measurements ---a
classical analog of this effect is the watched-pot paradox: a watched
pot never boils \cite{peres:AmJPhys:1980}---, although, generally
speaking, it could also be a system acted by some environment or even
the bare evolution of the system if the latter is not described by a
stationary state.
The inhibition of the evolution of a quantum system, though, was
already noted by von Neumann \cite{vonNeumann-bk:1932} and others
---an excellent account on the historical perspective of the QZE can
be found in \cite{pascazio:JPA:2008}.
From an experimental viewpoint, this effect was formerly detected by
Itano {\it et al.}\ \cite{itano:PRA:1990} considering the oscillations
of a two-level system, a modification suggested by Cook
\cite{cook:PhysScr:1988} of the original theoretical proposal.
Nonetheless,
the first experimental evidences with unstable systems, as originally
considered by Misra and Sudarshan, were observed later on by Raizen's
group \cite{raizen:Nature:1997,raizen:PRL:2001}.
Indeed, in the second experiment reported by this group in this regard
\cite{raizen:PRL:2001},
it was also shown the possibility to enhance the system decay by
considering measurements more spaced in time.
This is the so-called (quantum) anti-Zeno effect (AZE)
\cite{kaulakys:PRA:1997,luis:PRA:1998,kofman:Nature:2000}.

In the literature, it is common to introduce the QZE and the AZE
as antagonist, competing effects.
In this work, however, we study their manifestation within a unifying
framework, where they constitute the two limiting cases of a more
general process that we shall refer to as {\it quantum shuffling}.
To understand this concept, consider a series of von Neumann
measurements is performed on a quantum system, i.e., measurements such
that the outcome has a strong correlation with the measured quantity,
thus implying a high degree of certainty on the post-measured system
state.
This type of measurements provoke the system to collapse into any of the
pointer states of the measuring device, breaking the time coherence
that characterizes its unitary time-evolution.
In other words, the continuity in time between the pre and post-measurement
states of the system is irreversibly lost.
This loss takes place even when the distance (in time) between the
two states is so close that, in modulus, they look pretty much the
same, due to the corresponding loss of the phase accumulated with time
(which is a signature of the unitary time-evolution and therefore of
the possibility to revert the process in time).
Thus, consider the system is not stationary (regardless of the nature
of the source that leads to such non-stationarity), decaying
monotonically in time at a certain rate.
A series of measurements on this system will act similarly to
giving a deck of ordered cards a shuffle ---hence the name of quantum
shuffling---, affecting directly its coherence and modifying its
natural (unperturbed) decay time-scale.
As it is shown here, depending on the relative ratio between
the natural time-scale and the shufflingly-modified one,
the system decay can be either delayed or enhanced.
If the shuffling frequency (i.e., the amount of measurements per time
unit) is relatively low with respect to the system natural decay rate,
the pre and post-measurement system states will be very different.
This turns into a very fast decay due to the important lack of
correlation between both states.
Within the standard Zeno scenario, this enhancement of the decay
corresponds to AZE.
On the contrary, if the shuffling is relatively fast, the pre and
post-measurement states will be rather similar, for the system did
not have time enough to evolve importantly.
The decay is then slower, giving rise to QZE.
Now, as it is also shown, this inhibition of the system decay is only
apparent: the fast shuffling gives rise to an overall exponential decay
law at long times that makes the QZE to be sensitive to the total time
along which the system is monitored.
Thus, in the long term, one just finds out that the system evolution
displays features typical of Markovian processes \cite{breuer-bk:2002},
such as exponential decays (with relatively long characteristic
times) and time correlation functions with the form of a Markov chain
\cite{kloeden-bk:1992}.
As a consequence, if the natural system relaxation goes as a decreasing
power series with time (e.g., in systems with a regular system
dynamics), in the QZE regime it is found that the perturbed system
undergoes decays that fall below the natural decay after some time
due to the exponential decay induced by the short-spaced measurement
process.
This unexpected behavior is usually missed and therefore unexplored,
since in QZE scenarios it is more common to consider the mathematical
limit rather than the physical one.

In order to demonstrate the assertions mentioned above as clearly as
possible, here we have considered as a working model the free
evolution of a Gaussian wave packet.
When it is not perturbed by any measurement, this non-stationary system
undergoes a natural decay.
That is, this decay is not bound to effects linked to the action of
external potentials or surrounding environments, but only to the
bare wave-packet spreading as time proceeds.
From a time-independent perspective, this spreading is explained by the
continuum of frequencies or energies (plane waves) that contribute
coherently to the wave packet, which give rise to a non-stationary
evolution with time; from a time-dependent view, it is just a
diffraction effect associated with the initial {\it localization}
(spatial finiteness) of the wave packet.
In either case, this property together with the analyticity and
ubiquity of the model (it is a prototypical wave function describing
the initial state of atomic, molecular and optical systems) make of
the free Gaussian wave packet an ideal candidate to explore the Zeno
dynamics.
As it is shown, specific conditions for the occurrence of both QZE
and AZE are thus obtained in relation to the two mechanisms involved
in its dynamics: its translational motion and its intrinsic spreading,
which have been shown to rule the dynamics of quantum phenomena, such
as interference \cite{sanz:JPA:2008} or tunneling \cite{sanz:JPA:2011}.
More specifically, in order to detect QZE and AZE, the overlapping of
the wave function at two different times must be non-vanishing.
In this regard, therefore, if translation dominates the evolution, the
correlation function will decay relatively fast and none of them will
be observable.
The analytical results here obtained, properly adapted to other
contexts, may provide the physical insight necessary to understand
more complex processes described by the presence of external
interaction potentials or coupled environments
\cite{octavio:2011,tomas:2011}.
It is also worth stressing that, to some extent, the Zeno regimes
found keep a certain closeness with the three-time domain scenario
considered by Chiu, Sudarshan and Misra \cite{misra:PRD:1977}.

The work is organized as follows. The dynamics of a free wave
packet is introduced in Section~\ref{sec2}, in particular, the
(analytical) behavior of its associated time-dependent correlation
function. This will provide us with the basic elements to later on
establish the conditions leading to QZE or AZE once the shuffling
process induced by the succession of measurements will be introduced.
In particular, the quantum shuffling effect is analyzed in
Section~\ref{sec3} assuming the wave packet is acted by a series of
von Neumann measurements.
These measurements will be assumed to occur at equally spaced intervals
of time and their action on the system will be such that the
post-measurement state will always be equal to the initial one.
This could be the case, for example, when considering projections
(diffractions) through identical slits \cite{gonzalo:PRA:2011}.
Finally, in Section~\ref{sec4} we summarize the main conclusions
extracted from this work.


\section{Dynamics of a free wave packet}
\label{sec2}


\subsection{Characteristic time scales}
\label{sec21}

Consider the initial state of a quantum system is described in
configuration space by the Gaussian wave packet
\begin{equation}
 \Psi_0 (x) = A_0 e^{-(x-x_0)^2/4\sigma_0^2 + ip_0(x-x_0)/\hbar} ,
 \label{wfsanz080}
\end{equation}
where $A_0 = (2\pi\sigma_0^2)^{-1/4}$ is the normalization constant,
$x_0$ and $p_0$ are, respectively, the position and translational (or
propagation) momentum of its centroid, and $\sigma_0$ is its initial
spatial spreading.
The time-evolution of this wave function in free space ($V(x) = 0$)
is given \cite{sanz:JPA:2008} by
\begin{equation}
 \Psi_t (x) = A_t e^{-(x-x_t)^2/4\sigma_0\tilde{\sigma}_t
     + i\,p_0(x-x_t)/\hbar + iE_0 t/\hbar} .
 \label{wfsanz08}
\end{equation}
Here, $A_t = (2\pi\tilde{\sigma}_t^2)^{-1/4}$ is the
time-dependent normalization factor; $x_t = x_0 + v_0 t$ is the
time-dependent position of the wave packet centroid, with $v_0 =
p_0/m$ being its speed; $E_0 = p_0^2/2m$ is the average translational
energy, responsible for the time-dependent phase developed by the
wave packet as time proceeds; and
$\tilde\sigma_t = \sigma_0 \left[1 + (i\hbar
t/2m\sigma_0^2)\right]$, with $\sigma_t = |\tilde\sigma_t| =
\sigma_0\sqrt{1 + (\hbar t/2m\sigma_0^2)^2}$ being the time-dependent
spreading of the wave packet.
This spreading arises \cite{sanz:JPA:2008} from a type of internal or
intrinsic kinetic energy, which can be somehow quantified in terms of
the so-called spreading momentum, $p_s = \hbar/2\sigma_0$, an indicator
of how fast the wave packet will spread out.
This additional kinetic contribution becomes apparent when analyzing
the expectation value of the energy or average energy for the wave
packet,
\be
 \langle \hat{H} \rangle  =
  \frac{p_0^2}{2m} + \frac{p_s^2}{2m} ,
 \label{energy1wpa}
\ee
as well as in the variance,
\be
 \Delta E \equiv \sqrt{  \langle \hat{H}^2 \rangle
  - \langle \hat{H} \rangle^2 }
  = \sqrt{\frac{2p_s^2}{m}}\
   \sqrt{ \frac{p_0^2}{2m} + \frac{p_s^2}{4m}} .
 \label{deltaE}
\ee
(These two quantities are time-independent because of the commutation
between Hamiltonian operator, $\hat{H}$, and the time-evolution
operator, $\hat{U} = e^{it\hat{H}/\hbar}$.)
From a dynamical point of view, the implications of this term in
(\ref{energy1wpa}) are better understood through the real phase of
(\ref{wfsanz08}),
\begin{equation}
  S(x,t) =  p_0 (x - x_t)
   + \frac{\hbar t}{8m\sigma_0^2\sigma_t^2}\ (x-x_t)^2
   + E_0 t - \frac{\hbar}{2}\ (\tan)^{-1}
     \left( \frac{\hbar t}{2m\sigma_0^2} \right) .
 \label{Ssanz07}
\end{equation}
Putting aside the third and fourth terms in these expressions ---two
space-independent phases related to the propagation and normalization
in time, respectively---, we observe that the first term is a
classical-like phase associated with the propagation itself of the wave
packet, while the second one is a purely quantum-mechanical phase
associated with its spreading motion.
Correspondingly, each one of these two motions leads to the two
energy contributions that we find in (\ref{energy1wpa}).

By inspecting the functional dependence of $\sigma_t$ on time, a
characteristic time scale can be defined, namely
$\tau \equiv 2m\sigma_0^2/\hbar$.
This time scale is associated with the relative spreading of the wave
packet, allowing us to distinguish three dynamical regimes in its
evolution depending on the ratio between $t$ and $\tau$
\cite{sanz:AJP:2011}:
\begin{enumerate}[(i)]
 \item The very-short-time or Ehrenfest-Huygens regime,
  $t \ll\!\!< \tau$, where the wave packet remains almost spreadless:
  $\sigma_t \approx \sigma_0$.
 \item The short-time or Fresnel regime, $t \ll \tau$, where the
  spreading increases nearly quadratically with time:
  $\sigma_t \approx \sigma_0 + (\hbar^2/8m^2\sigma_0^3) t^2$.
 \item The long-time or Fraunhofer regime, $t \gg \tau$, where the
  Gaussian wave packet spreads linearly with time: $\sigma_t \approx
  (\hbar/2m\sigma_0) t$.
\end{enumerate}
By means of $\tau$ we can thus characterize the dynamics
of the Gaussian wave packet, although similar time scales could also
be found in the case of more general wave packets provided that
we have at hand their time-dependent trend ---this is in correspondence
with the time-domains determined by Chiu, Sudarshan and Misra
for unstable systems \cite{misra:PRD:1977}.
Keeping this in mind, consider the probability density associated
with (\ref{wfsanz08}),
\begin{equation}
 |\Psi_t(x)|^2 = \frac{1}{\sqrt{2\pi\sigma_t^2}}\
    e^{-(x-x_t)^2/2\sigma_t^2} .
 \label{probsanz07}
\end{equation}
Case (i) is not interesting, because it essentially implies no
evolution in time.
So, let us focus directly on case (ii), for which (\ref{probsanz07})
reads as
\begin{equation}
 |\Psi_t(x)|^2 \approx \frac{1}{\sqrt{2\pi\sigma_0^2}}
  \left[ 1 - \left(\frac{\hbar^2}{8m^2\sigma_0^4}\right) t^2 \right]
    e^{-(x-x_t)^2/2\sigma_0^2} ,
 \label{probsanz07a}
\end{equation}
where the time-dependent factor in the argument of the exponential
can be neglected without loss of generality (the exponential of
such an argument is nearly one).
According to (\ref{probsanz07a}), the initial falloff of the
probability density is parabolic and therefore susceptible to display
QZE if a series of measurement is carried out at regular intervals of
time \cite{dorlas:SPIE:2004,helen:JPCM:2011} provided that
these time intervals are, at least, $\Delta t \lesssim \tau$ (later on,
in Section~\ref{sec3}, another characteristic time scale, namely the
Zeno time, will also be introduced).
For longer time scales (case (iii)),
\begin{equation}
 |\Psi_t(x)|^2 \approx \sqrt{\frac{2m^2\sigma_0^2}{\pi\hbar^2 t^2}}\
   e^{- (2m^2\sigma_0^2/\hbar^2) (x-x_t)^2/t^2}
 = \frac{1}{\sqrt{2\pi\sigma_0^2}}\frac{\tau}{t}\
   e^{- (\tau/t)^2 (x-x_t)^2/2\sigma_0^2} .
 \label{probsanz07b}
\end{equation}
Accordingly, for distances such that the ratio $(x-x_t)/t$ remains
constant with time (remember that the spreading is now linear with
time), the probability density will decay like $t^{-1}$, leading to
observe AZE instead of QZE.
As it will be shown below in
more detail, note that the effect of introducing $N$ measurements
is equivalent (regardless of constants) to having $t^{-N}$, which
goes rapidly to zero.


\subsection{Correlation functions and survival probabilities}
\label{sec22}

Now we shall focus on the quantity central to the discussion in this
work: quantum correlation function at two different times.
Thus, let us consider $|\Psi_{t_1}\rangle$ generically denotes the state of
a quantum system at a time $t_1$.
The unitary time-evolution of this state from $t_1$ to $t_2$ (with
$t_2 > t_1$) is accounted for the formal solution of the time-dependent
Schr\"odinger equation
\be
 |\Psi_{t_2}\rangle = \hat{U}(t_2,t_1) |\Psi_{t_1}\rangle
  = e^{-i\hat{H}(t_2-t_1)/\hbar} |\Psi_{t_1}\rangle .
 \label{timev}
\ee
The quantum time-correlation function is defined as
\be
 C(t_2,t_1) \equiv \langle \Psi_{t_1} | \Psi_{t_2} \rangle ,
 \label{corrinit}
\ee
measuring the correlation existing between the system states at $t_2$
and $t_1$, or, equivalently, after a time $t = t_2 - t_1$ has elapsed
(since $t_1$).
This second notion also allows us to rewrite (\ref{corrinit}) as the
correlation function between the wave function at a time $t = t_2 -
t_1$ and the initial wave function ($t=0$), as it follows from
\be
 C(t_2,t_1) = \langle \Psi_{t_1} | \Psi_{t_2} \rangle
  = \langle \Psi_0 | \hat{U}^+(t_1) \hat{U}(t_2) | \Psi_0 \rangle
  = \langle \Psi_0 | \hat{U}(t_2 - t_1) | \Psi_0 \rangle
  = \langle \Psi_0 | \Psi_t \rangle = C(t) .
\ee
Another related quantity of interest here is the survival probability,
\be
 P(t_2,t_1) \equiv | \langle \Psi_{t_1} | \Psi_{t_2} \rangle |^2
  = | \langle \Psi_0 | \Psi_t \rangle |^2 = P(t) .
 \label{defps}
\ee
This quantity indicates how much of the wave
function at $t_1$ still survives at $t_2$ (in both norm and phase) or,
equivalently, how much of the initial wave function (also, in norm and
phase) survives at a later time $t = t_2 - t_1$.
From now on, concerning the Zeno scenario, we are going to work
assuming the second approach, although it can be shown that both
are equivalent (see Appendix~\ref{appA}).
Taking this into account together with the general solution
(\ref{timev}), the short-time behavior of $P(t)$ can be readily found,
\be
 P(t) \approx
  | \langle \Psi_0 | \left( 1 - \frac{i\hat{H}t}{\hbar}
  - \frac{\hat{H}^2 t^2}{2\hbar^2} \right) | \Psi_0 \rangle |^2
  = 1 - \frac{(\Delta E)^2 t^2}{\hbar^2} ,
 \label{survapprox}
\ee
after considering a series expansion up to the second order in $t$ as
well as the normalization of $\Psi_0$.

In our case, in particular, the fact that the wave packet spreads along
time indicates that the quantum system becomes more delocalized, this
making the corresponding correlation function to decay.
This can be formally seen by computing the correlation function
associated with (\ref{wfsanz08}), which reads as
\be
 C(t) = \sqrt{\frac{2\sigma_0}{\sigma_0 + \tilde{\sigma}_t}}\
  e^{- E_0t^2/2m\sigma_0(\sigma_0 + \tilde{\sigma}_t) - iEt/\hbar}
  = \left[ 1 + \left( \frac{t}{2\tau} \right)^2 \right]^{-1/4}
  e^{- E_0t^2/4m\sigma_0^2[1 + (t/2\tau)^2] + i\delta_t} ,
 \label{corr}
\ee
with
\be
 \delta_t = \frac{1}{1 + (t/2\tau)^2}\frac{E_0t}{\hbar}
  - \frac{1}{2}\ (\tan)^{-1} \left( \frac{t}{2\tau} \right) .
\ee
The exponential in (\ref{corr}) only depends on the initial
momentum associated with the wave packet centroid, but not on its
initial position. This is a key point, for the loss of correlation in
a wave function displaying a translation faster than its spreading
rate will mainly arise from the lack of spacial overlapping
between its values at $t_1$ and $t_2$ (or, equivalently, at $t_0$
and $t$), rather than to the distortion of its shape (and
accumulation of phase).
However, a relatively slow translational motion will imply that the
loss of correlation is mainly due to the wave-packet spreading.
In this regard, note how the spreading acts as a sort of intrinsic
instability, which is not related at all with the action of an external
potential or a coupling to a surrounding environment, but that only
comes from the fact that the state describing the system is not
stationary (i.e., an energy eigenstate of the Hamiltonian, as it would
be the case of a plane wave).

Two scenarios can be thus envisaged to elucidate the mechanisms leading
to the natural loss of correlation in a quantum system (at
this stage, no measurement is assumed).
First, consider $p_0 = 0$, i.e., the wave packet only spreads
with time, first quadratically and then linearly after the boosting
phase \cite{sanz:cpl:2007}, as seen in Section~\ref{sec21}.
In this case the correlation function (\ref{corr}) reads as
\be
 C(t) = \left[ 1 + (t/2\tau)^2 \right]^{-1/4} e^{i\varphi_t} ,
 \label{corr1}
\ee
with
\be
 \varphi_t = - \frac{1}{2}\ (\tan)^{-1} \left( \frac{t}{2\tau} \right) ,
 \label{phase1}
\ee
and the corresponding survival probability (\ref{defps}) as
\be
 P(t) = \frac{1}{\sqrt{1 + (t/2\tau)^2}} .
 \label{surv1}
\ee
For short times (case (ii) above), (\ref{surv1}) becomes
\be
 P(t) \approx 1 - \frac{t^2}{8\tau^2} .
 \label{surv1short}
\ee
The functional form displayed by (\ref{surv1short}) is the typical
quadratic-like decay expected for any general quantum state, as it can
easily be seen by substituting (\ref{deltaE}) into the right-hand side
of the second equality of (\ref{survapprox}) with $p_0 = 0$.
Conversely, at very long times,
\be
 P(t) \approx \frac{2\tau}{t} ,
 \label{surv1long}
\ee
i.e., the survival probability decreases monotonically as
$t^{-1}$, in correspondence with the result found above for the
asymptotic behavior of the probability density. As time evolves,
the global phase of the correlation function goes from a linear
dependence with time ($\varphi_t \approx - t/4\tau$) to an
asymptotic constant value, $\varphi_\infty = - \pi/4$.
Its value thus remains bound at any time between 0 and
$\varphi_\infty$.

In the more general case of nonzero translational motion for the wave
packet ($p_0 \ne 0$), the survival probability is given by
\be
 P(t) = \frac{1}{\sqrt{1 + (t/2\tau)^2}}\
  e^{- E_0t^2/2m\sigma_0^2[1 + (t/2\tau)^2]} .
 \label{surv}
\ee
In the short-time limit, this expression reads as
\be
 P(t) \approx \left(1 - \frac{t^2}{8\tau^2}\right)
  e^{- E_0t^2/2m\sigma_0^2} .
 \label{surv2short}
\ee
which remarkably stresses the two aforementioned
mechanisms competing for the loss of the system correlation:
the spreading of the wave packet and its translational motion.
This means that if the translational motion is faster than the
spreading rate, the wave functions at $t_0$ and $t$ will not overlap,
and $P(t)$ will vanish very fast.
On the contrary, if the translational motion is relatively slow, the
overlapping will be relevant and the decay of $P(t)$ will go
quadratically with time. In order to express the relationship
between spreading and translation more explicitly, (\ref{surv2short})
can be expressed in terms of $p_0$ and $p_s$, i.e.,
\be
 P(t) \approx \left(1 - \frac{t^2}{8\tau^2}\right)
   e^{- 2 (p_0/p_s)^2 (t^2/8\tau^2)} .
 \label{surv2short1}
\ee
Thus, if the translational and spreading motions are such that
\be
 \frac{p_0}{p_s} \ll \frac{2\tau}{t}
\ee
(actually, it is enough that $p_0/p_s \lesssim 1/\sqrt{2}$, since
$t^2/8\tau^2$ is already relatively small),
then
\be
 P(t) \approx 1 - \left[ 1 + 2\left(\frac{p_0}{p_s}\right)^2\right]
  \frac{t^2}{8\tau^2} ,
 \label{surv2short2}
\ee
which again decays quadratically with time. Otherwise, the
decrease of $P(t)$ will be too fast to observe either QZE or AZE
(see below). Regarding the long-time regime, we find
\be
 P(t) \approx \frac{2\tau}{t}\ e^{- 2E\tau^2/m\sigma_0^2}
  = \frac{2\tau}{t}\ e^{- (p_0/p_s)^2} ,
 \label{surv2long}
\ee
which displays the same decay law ($t^{-1}$) as in the case $p_0 =
0$, since the argument of the exponential function becomes
constant. Regarding the phase $\delta_t$, it should be mentioned
that at short times it depends linearly with time, increasing or
decreasing depending on which mechanism (translation or spreading)
is stronger. However, at longer times it approaches asymptotically
(also like $t^{-1}$) the value $\varphi_\infty$ regardless of
which mechanism is the dominant one.


\section{Quantum Zeno effect and projection operations}
\label{sec3}

In the standard QZE scenario, a series of von Neumann
measurements are performed on the system at regular intervals of
time $\Delta t$.
Between two any consecutive measurements the system follows a unitary
time-evolution according to (\ref{timev}), while each time a
measurement takes place (at times $t = n \Delta t$, with $n = 1, 2,
\ldots$) the unitarity of the process breaks down and the system
quantum state ``collapses'' into one of the pointer states of the
measuring device.
With this scheme in mind, consider the pointer states are equal to the
system initial state ---in the case we are analyzing here, this type
of measurements could consist, for example, of a series of diffractions
produced by slits with similar transmission properties to the one that
generated the initial wave function \cite{gonzalo:PRA:2011}.
Thus, after the first measurement the system state will be
\be
 |\Psi_{t=\Delta t}\rangle =
  |\Psi_0\rangle \langle \Psi_0 | \Psi_{\Delta t} \rangle ,
\ee
which coincides with the initial state, although its amplitude is
decreased by a factor $\langle \Psi_0 | \Psi_{\Delta t} \rangle$.
Each new measurement will therefore add a multiplying factor
$|\langle \Psi_0 | \Psi_{\Delta t} \rangle|^2$ in the survival
probability, which implies that it will read as
\be
 P_n(t) = \left[ \mathcal{P}_{\Delta t}^{(0)} \right]^n
  |\langle \Psi_0 | \Psi_{t-n\Delta t} \rangle|^2
 \label{survn}
\ee
after $n$ measurements, where $\mathcal{P}_{\Delta t}^{(0)} \equiv
|\langle \Psi_0 | \Psi_{\Delta t} \rangle|^2$.
For $\Delta t$ sufficiently small, $\mathcal{P}_{\Delta t}^{(0)}$
acquires the form of (\ref{survapprox}),
\be
 \mathcal{P}_{\Delta t}^{(0)} \approx
  1 - \frac{(\Delta E)^2 (\Delta t)^2}{\hbar^2} ,
 \label{uncert}
\ee
from which another characteristic time arises, namely the {\it Zeno
time} \cite{pascazio:JPA:2008}, defined as
\be
 \tau_Z \equiv \frac{\hbar}{\Delta E} .
 \label{zenotimedef}
\ee
In Section~\ref{sec21}, different stages in the natural evolution of
the quantum system were distinguished given the ratio between $t$ and
the time scale $\tau$.
The new time scale provided by $\tau_Z$ also allows us to distinguish
between two types of dynamical behavior.
For measurements performed at intervals such that $\Delta t \ll \tau_Z$,
(\ref{uncert}) holds and the decay of the perturbed correlation function
will be relatively slow with respect to the total time the system is
monitored.
Traditionally, this defines the Zeno regime, where the decay of the
correlation function is said to be inhibited due to the measurements
performed on the system.
On the contrary, as $\Delta t$ becomes closer to $\tau_Z$, (\ref{uncert})
does not hold anymore and the decay of the correlation function
becomes faster than the unperturbed one for finite $t$.

In the case of a free Gaussian wave packet with $p_0 = 0$ (the system
dynamics is only ruled by the wave packet spreading),
substituting (\ref{deltaE}) into (\ref{zenotimedef}) the Zeno time can
be expressed as
\be
 \tau_Z = 2\sqrt{2}\ \! \tau ,
\ee
which is nearly three times larger than $\tau$.
According to the standard scenario, provided that $\Delta t$ is smaller
than $\tau_Z$, one should observe QZE.
However, the characteristic time $\tau$ also plays a key role: as shown
below, QZE is observable provided that measurements are performed at
time intervals much shorter than the time scales ruling the wave-packet
linear spreading regime.
Otherwise, only AZE will be observed.
If now we consider the more general case, where the free wave packet
has an initial momentum ($p_0 \ne 0$), a more stringent condition is
obtained.
According to (\ref{surv2short2}) ---or, equivalently, substituting
(\ref{deltaE}) into definition (\ref{zenotimedef})---, we find
\be
 \tau_Z = \frac{2\sqrt{2}\ \! \tau}{\sqrt{1 + 2(p_0/p_s)^2}} ,
\ee
which implies that, in order to observe QZE, the time intervals
$\Delta t$ between two consecutive measurements have to be even shorter
(apart from the fact that the condition $p_0/p_s \lesssim 1/\sqrt{2}$
should also be satisfied).
This condition ensures that the wave function at $t$ still has an
important overlap with its value at $t_0$.

\begin{figure}[!t]
 \begin{center}
 \includegraphics[width=14cm]{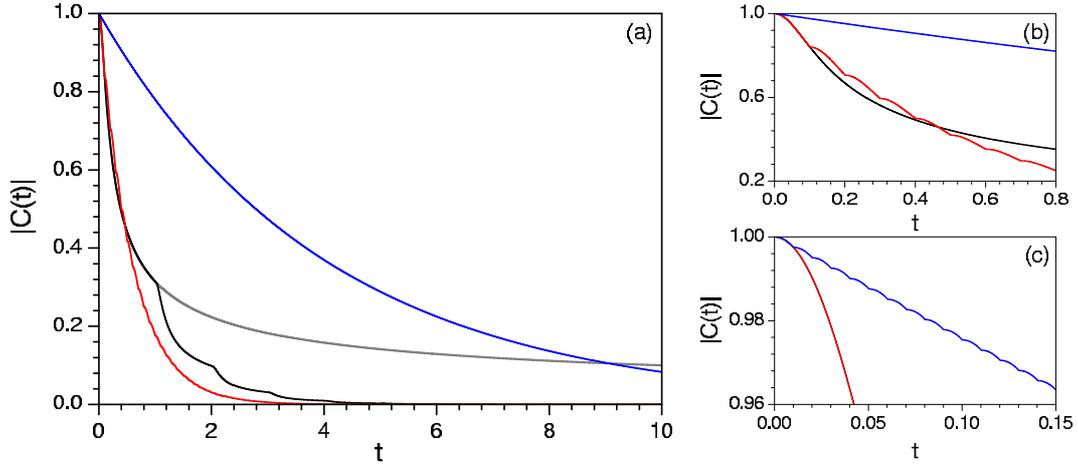}
 \caption{(a) Modulus of the time correlation function, $|C(t)|$, for
  the unperturbed system (gray line) and three different cases with
  measurements performed at: $\Delta t_1 = 10^4\ \! \delta t = 1$ (black),
  $\Delta t_2 = 10^3\ \! \delta t = 0.1$ (red), and
  $\Delta t_3 = 10^2\ \! \delta t = 0.01$ (blue),
  with $\delta t = 10^{-4}$
  being the time-step considered in the simulation.
  (b) and (c) are enlargements of part (a) for times of the order of
  $\tau_Z$ and $\tau$, respectively.
  In the calculations, $m = 0.1$, $\sigma_0 = 0.5$ and $p_0 = 0$,
  which render $\tau = 0.05$ and $\tau_Z = 0.14$ (see text for
  details).}
 \label{fig1}
 \end{center}
\end{figure}

With the tools developed so far, let us now have a closer
look at the QZE and AZE dynamics.
Typically, these effects are assumed to be quite the opposite.
However, we show they constitute the two limits of the
aforementioned quantum shuffling process.
For simplicity and without loss of generality, instead of considering
the survival probability, in Fig.~\ref{fig1}(a) we have plotted the
modulus of the time correlation function, $|C(t)|$, against time to
monitor the natural (unperturbed) evolution of the wave packet (gray
curve) and three cases where measurements have been performed at
different time intervals $\Delta t$.
These intervals have been chosen proportional to the time-step
$\delta t$ ($= 10^{-4}$ time units) used in the numerical simulation:
$\Delta t_1 = 10^4\ \! \delta t = 1$ (black),
$\Delta t_2 = 10^3\ \! \delta t = 0.1$ (red),
and $\Delta t_3 = 10^2\ \! \delta t = 0.01$ (blue).
Regarding other parameters, we have used
$m = 0.1$, $\sigma_0 = 0.5$ and $p_0 = 0$, which make $\tau =
0.05$ and $\tau_Z \approx 0.14$.
The three color curves displayed in Fig.~\ref{fig1}(a), which show
the action of a set of measurements on the quantum system, behave in a
similar fashion: they are piecewise functions, each piece being
identical to the corresponding one between $t = 0$ and $t = \Delta t$,
i.e., to $\mathcal{C}_{\Delta t}^{(0)} \equiv
\sqrt{\mathcal{P}_{\Delta t}^{(0)}}$.
These curves allow us to illustrate the quantum shuffling process in
three time regimes which depend on the relationship between $\tau$,
$\tau_Z$ and $\Delta t$:
\begin{enumerate}[(a)]
 \item For $\tau < \tau_Z \le \Delta t$, the correlation function
  (see the black curve in Fig.~\ref{fig1}(a))
  is clearly out of the quadratic-like time domain,
  $\mathcal{C}_{\Delta t}^{(0)}$ is convex and therefore the perturbed
  correlation function {\it always} goes to zero much faster than the
  natural decay law (gray curve).
  This is what we call {\it pure} AZE, for the correlation function is
  always decaying below the unperturbed function.

 \item For $\tau \le \Delta t \le \tau_Z$, according to the literature
  one should observe QZE.
  However, this is not exactly the case.
  Between $\tau$ and $\tau_Z$, the correlation function (\ref{surv1})
  displays an inflection point at $\tau_{\rm inflx} = \sqrt{2}\ \! \tau
  \approx 0.071$, changing from convex to concave.
  Thus, for $\Delta t$ between $\tau_{\rm inflx}$ and $\tau_Z$,
  pure AZE is still found due to the convexity of the time correlation
  function.
  Now, if $\tau \le \Delta t \le \tau_{\rm inflx}$, the initial falloff
  of the perturbed correlation function is slower,
  $\mathcal{C}_{\Delta t}^{(0)}$ becomes concave and the overall decay
  gets slower than that associated with the unperturbed correlation
  function (see red curve in Fig.~\ref{fig1}(a)).
  This lasts out for some time, after which the perturbed correlation
  function falls below the unperturbed one (see red curve in
  Fig.~\ref{fig1}(b)).
  It is worth stressing here how the decay is indeed faster than in the
  case of pure AZE, with the quantum shuffling making the perturbed
  correlation function to acquire a seemingly exponential-like shape.

 \item For $\Delta t < \tau$, the wave packet is well inside the
  region where the wave function decay is quadratic-like (and concave)
  and therefore the quantum shuffling produces decays much slower than
  those observed in the unperturbed correlation function as $\Delta t$
  decreases (see blue curve in Fig.~\ref{fig1}(c)).
  This is commonly known as QZE.
  Now, this inhibition of the decay is only apparent; if one considers
  longer times (see blue curve in Fig.~\ref{fig1}(a)), the correlation
  function is essentially a decreasing exponential, which eventually
  leads the (perturbed) system to decay to zero earlier than its
  unperturbed counterpart.
  As it will be shown below, these exponential decays can be justified
  in terms of a sort of Markovianity induced by the shuffling process
  on the system evolution.
\end{enumerate}

\begin{figure}[!t]
 \begin{center}
 \includegraphics[width=14cm]{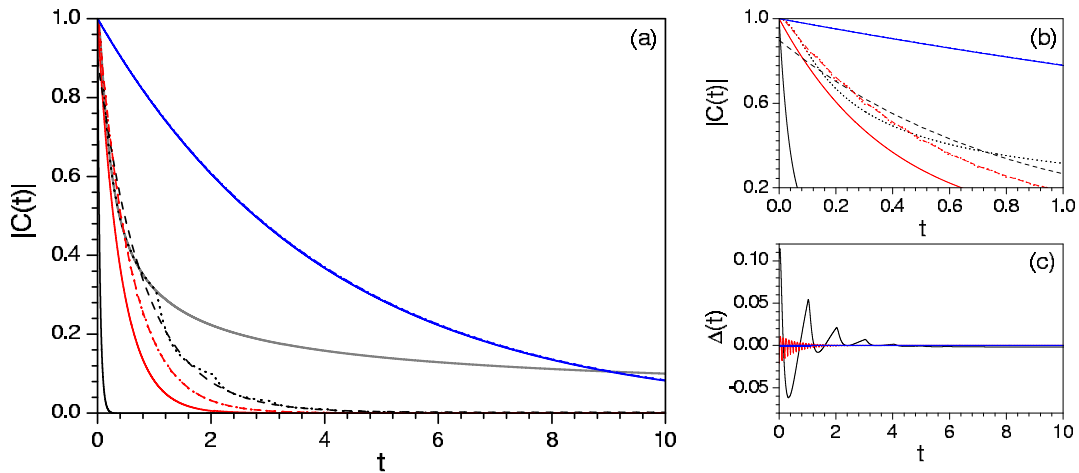}
 \caption{(a) Same as Fig.~\ref{fig1}(a), but showing the envelope
  (\ref{chain2}) superimposed to the corresponding perturbed decay
  functions: $\Delta t_1 = 1$ (black), $\Delta t_2 = 0.1$ (red) and
  $\Delta t_3 = 0.01$ (blue).
  In the figure, the different types of line denote the modulus of the
  time correlation function, $|C(t)|$, obtained from: the simulation
  (dotted), the theoretical estimation (\ref{chain2}) (solid), and the
  fitting to a pure exponential function (dashed);
  to compare with, the unperturbed
  correlation function is also displayed with gray line in panel (a).
  The decay rates arising from the theoretical
  estimation are $\gamma'_{1,\rm est} = 25$ (black),
  $\gamma'_{2,\rm est} = 2.5$ (red) and $\gamma'_{3,\rm est} = 0.25$
  (blue), while those obtained from the fitting are
  $\gamma'_{1,\rm fit} = 1.224$ (black),
  $\gamma'_{2,\rm fit} = 1.734$ (red) and
  $\gamma'_{3,\rm fit} = 0.249$ (blue).
  (b) Enlargement of part (a) in the time interval between $t = 0$ and
  $t = 1$.
  (c) Plot of the difference $\Delta(t)$ between the estimated
  envelope, $\mathcal{C}_{\Delta t}(t)$, and the fitted envelope,
  in part (a).}
 \label{fig2}
 \end{center}
\end{figure}

In order to better understand the subtleties behind the quantum
shuffling dynamics (and therefore the QZE and the AZE), let us focus
only on the overall prefactor that appears in (\ref{survn}), which in
the short-time regime can be written as
\be
 \mathcal{P}_{\Delta t}^{(n)} \equiv
  \left[ \mathcal{P}_{\Delta t}^{(0)} \right]^n
  \approx \left[ 1 - \frac{(\Delta t)^2}{\tau_Z^2} \right]^n .
 \label{pn}
\ee
This is a
discrete function of $n$, the number of measurements performed up
to $t_n \equiv n \Delta t$, the time at which the $n$-th measurement
is carried out.
In the limit $n \to \infty$, (\ref{pn}) becomes
\be
 \mathcal{P}_{\Delta t}^{(\infty)} (t_n) \approx
  e^{-\gamma_{\Delta t} t_n} ,
 \label{chain}
\ee
with the decay rate being
\be
 \gamma_{\Delta t} \equiv \frac{\Delta t}{\tau_Z^2} ,
 \label{decayrate}
\ee
as also noted in \cite{pascazio:JPA:2008}.
This rate defines another characteristic time, $\tau_{\Delta t} \equiv
\gamma_{\Delta t}^{-1}$, associated with the falloff of the continuous
form of (\ref{chain}),
\be
 \mathcal{P}_{\Delta t}(t) = e^{-\gamma_{\Delta t} t}
 \label{chain2}
\ee
(note that this function passes through all the points $t_n$ upon
which (\ref{chain}) is evaluated).
In Fig.~\ref{fig2} we show a comparative
analysis between the correlation functions $|C(t)|$ displayed in
Fig.~\ref{fig1} and their respective envelopes, given by
$\mathcal{C}_{\Delta t}(t) \equiv \sqrt{\mathcal{P}_{\Delta t} (t)}$;
the former are denoted with dotted
line and the latter with solid line of the same color (again, the gray
solid line represents the unperturbed correlation function).
The values for the estimated decay rates, given by
$\gamma'_{\Delta t} = \gamma_{\Delta t}/2$ for the curves represented,
are: $\gamma'_1=25$ (black), $\gamma'_2=2.5$ (red) and $\gamma'_3=0.25$
(blue).
As it can be seen, the agreement between the correlation function and
its envelope $\mathcal{C}_{\Delta t}(t)$ becomes better as $\Delta t$
decreases (see Figs.~\ref{fig2}(a) and (b)), which is in virtue of the
approximation considered in (\ref{pn}) ---as
$\Delta t$ increases the behavior of the envelope (\ref{chain2}) will
diverge more remarkably with respect to the trend displayed by $P_n(t)$,
whereas both will converge as $\Delta t$ becomes smaller.
Thus, while for long intervals $\Delta t$ between consecutive measurements
the envelope deviates importantly from the associated correlation
function (see black dotted and solid lines), as $\Delta t$ becomes
smaller the difference between both curves reduces importantly and
the relaxation takes longer times (of the order of $\tau_{\Delta t}$).
Nevertheless, for larger values of $\Delta t$ one can still perform a
fitting of the correlation function to a decaying exponential function,
$\mathcal{C}_{\Delta t, \rm fit}^{(\infty)}(t) = e^{-\gamma'_{\rm fit}t}$,
which renders a qualitatively good overall agreement, as can be seen
from the corresponding dashed lines in Fig.~\ref{fig2}(a).
Note that the decay rates obtained from this fitting are closer to the
falloff observed for the corresponding correlations functions
($\gamma'_{1,\rm fit} = 1.224$, $\gamma'_{2,\rm fit} = 1.734$, and
$\gamma'_{3,\rm fit} = 0.249$), converging to the estimated value
$\gamma'$ as $\Delta t$ decreases ($\gamma'_{3,\rm fit} \approx
\gamma'_3$).

From the previous discussion, the idea of a series of sequential
measurement acting as a shuffling process, wiping out any memory of the
system past history, arises in a natural way.
Hence, as $\Delta t$ becomes smaller, $|C(t)|$ becomes
closer to its envelope $\mathcal{C}_{\Delta t}(t)$ and
therefore to an exponential decay law.
Conversely, for larger values of $\Delta t$, the system keeps memory
of its past evolution for relatively longer periods of time (between
two consecutive measures), this leading to larger discrepancies between
the correlation function and (\ref{chain2}).
Taking into account this point of view and getting back to
(\ref{survn}), one can notice a remarkable resemblance between
this expression and a Markov chain of independent processes
\cite{kloeden-bk:1992} (which
is also inferred from (\ref{chain})): the state after one measurement
only depends on the state before it, but not on the previous history
or sequence until this state is reached.
That is, between any two consecutive measurements
we have a precise knowledge of the probability to find the system
in a certain time-dependent state, while, after a measurement, we
loose any memory on that. Thus, as $\Delta t$ becomes smaller, the
process becomes fully Markovian, with the time correlation function
approaching the typical exponential-like decreasing behavior
characteristic of this type of processes.
On the contrary, as $\Delta t$ increases, the memory on the past
history is kept for longer times, this turning the system evolution
into non-Markovian, which loses gradually the smooth exponential-like
decay behavior.
Only when a measurement is carried out such memory is suddenly
removed, which is the cause of the faster (sudden) decays
observed in black curve of Fig.~\ref{fig1}(a).

The transition from the non-Markovian to the Markovian regime can be
somehow quantified by monitoring along time the distance between the
estimated envelope, $\mathcal{C}_{\Delta t}(t)$, and the fitted
envelope, $\mathcal{C}_{\Delta t, \rm fit}(t)$,
\be
 \Delta(t) \equiv
  \mathcal{C}_{\Delta t}(t) - \mathcal{C}_{\Delta t, \rm fit}(t) ,
\ee
which is plotted in Fig.~\ref{fig2}(c) for the three cases of
$\Delta t$ considered.
Thus, as $\Delta t$ becomes smaller, we approach an exponential decay
law and $\Delta(t)$ goes to zero for any time (see blue curve in the
figure), this being the signature of Markovianity.
On the contrary, if the time evolution is not Markovian, as time
increases and the system keeps memory for longer times, the value of
$\Delta(t)$ displays important deviations from zero (see black and
red curves in the figure).
These deviations mainly concentrate on the short and medium term
dynamics, where values are relatively large to be remarkable.
Nonetheless, analogously, one could also display the relative ratio
between the two correlation functions, which would indicate or not
the trend toward Markovianity in the long-time (asymptotic) regime.


\section{Conclusions}
\label{sec4}

By assuming that the QZE inhibits the evolution of an unstable quantum
system, one might also be tempted to think that its coherence is also
preserved, while the AZE would lead to the opposite effect in its way
through faster system decays.
In order to better understand these effects, here we have focused
directly on the bare system, i.e., no external potentials or
surrounding environments acting on the quantum system have been
assumed.
This has allowed us to elucidate the conditions under which such
effects take place in relation to the intrinsic time-scales
characterizing the isolated system, which have been shown to play
an important role.
Furthermore, by means of this analysis, we have also shown that both
QZE and AZE are indeed two instances of a more general effect, namely
a {\it quantum shuffling process}, which eventually leads the system to
display a Markovian-like evolution and its correlation function to
follow an exponential decay law as the interval between measurements
decreases.
Within this scenario, the QZE dynamics can be regarded as a regime
where any trace of knowledge on the initial system state is lost due
to a rapid shuffling, while in the long-time regime the correlation
function would fall to zero faster than the unperturbed one.
The apparent contradiction with the traditional no-evolution scenario
can be explained very easily: Since one often cares only about the
short-time dynamics, the long-time dynamics is usually completely
neglected.
In other words, the time during which the system dynamics is usually
studied is relatively small compared to the Markovian time-scale
induced by the continuous measurement process and, therefore, one
assumes nearly stationary dynamics.



\section*{Acknowledgements}

The support from the Ministerio de Ciencia e Innovaci\'{o}n (Spain)
through Projects FIS2010-18132, FIS2010-22082, CSD2009-00038; from
Comunidad Aut\'onoma de Madrid through Grant No.\ S-2009/MAT/1467;
and from the COST Action MP1006 ({\it Fundamental Problems in Quantum
Physics}) is acknowledged.
A. S. Sanz also thanks the Ministerio de Ciencia e Innovaci\'{o}n for
a ``Ram\'on y Cajal'' Research Fellowship.


\appendix

\section{Alternative Zeno scenario}
\label{appA}

In the Zeno scenario considered above, establishing a direct analogy
with Zeno's arrow, the wave packet plays the role of a quantum
arrow, but with the particularity that this arrow slows down until
its evolution is frozen by means of a series of measurements. This
is the scenario traditionally considered \cite{pascazio:JPA:2008}.
However, a more direct analogy with Zeno's arrow paradox can be
established if it is assumed that the wave packet is always in
motion and the measurements are just like photographs indicating the
particular instant from which the correlation has to be computed
\cite{tomas:2011}. Because of the actions on
the wave packet in relation to what a measurement is considered in
each, we can call these two situations as:
\begin{enumerate}[(a)]
 \item The {\it stopping-arrow scenario}, where the wave packet is
  ``collapsed'' or ``stopped'' after each measurement.
 \item The {\it steady-arrow scenario}, where the wave packet
  time-evolution never stops, but the computation of the correlation
  function is reset after each (photograph-like) measurement ---like
  in a stop-motion or stop-action movie.
\end{enumerate}

It can be shown that both scenarios are equivalent with the exception
of a lost time-dependent phase in the latter.
In order to prove this statement, let us start by considering the
wave function is now left to freely evolve in time.
Following the idea behind this scenario, the survival probability is
monitored in time by computing the overlapping of the wave function
at a time $t$ with its value at successive times $t_0$, $t_1$,
$t_2$, with $t_n = n \Delta t$, i.e.,
\be
 \langle \Psi_{t_{n-1}} | \Psi_t \rangle ,
\ee
where $t_{n-1} \le t < t_n$.
Note here the direct analogy with Zeno's arrow, where at each
instant we are observing the arrow steady at a different space position,
but without freezing its motion.
Taking this into account, for $0 \le t < t_1$, we have
\be
 P(t) = |\langle \Psi_0 | \Psi_t \rangle|^2 .
\ee
Now, if $t_1 \le t < t_2$,
\be
 P(t) = \alpha_1 |\langle \Psi_{t_1} | \Psi_t \rangle|^2 ,
\ee
while the wave function for the same interval will be
\be
 |\Psi_t\rangle = \sqrt{\alpha_1} |\Psi_t\rangle ,
\ee
i.e., the evolved wave function, but with a prefactor which ensures
the matching of the different of $P(t)$ before and after $t = t_1$.
Following the same argumentation, after the $n-1$ measurement,
\be
 P(t) = \left( \Pi_{k=1}^{n-1} \alpha_k \right)
  |\langle \Psi_{t_{n-1}} | \Psi_t \rangle|^2 ,
 \label{photo}
\ee
with $t_{n-1} \le t \le t_n$.

As it can be seen, by means of this procedure the wave function is
never altered (which always keeps evolving according to the
Schr\"odinger equation), but only its relative amplitude.
So, the key issue here is the attenuation factor $\alpha$, which can
be evaluated as follows.
For $t_{n-1} \le t < t_n$, we note that
\be
 \langle \Psi_{t_{n-1}} | \Psi_t \rangle
  = \langle \Psi_0 | e^{i\hat{H}(n-1)\Delta t/\hbar}
    e^{-i\hat{H}t/\hbar} | \Psi_0 \rangle
  = \langle \Psi_0 | e^{i\hat{H}[t - (n-1)\Delta t]/\hbar} | \Psi_0 \rangle .
\ee
Therefore, the attenuation factor for the interval $t_{n-1} \le t \le t_n$
should come from the overlapping of the wave function at the times when
the two previous measurements were performed, i.e.,
\be
 \alpha_{n-1} = |\langle \Psi_{t_{n-2}} |
  \Psi_{t_{n-1}} \rangle |^2
 = |\langle \Psi_0 | e^{i\hat{H}\Delta t/\hbar} | \Psi_0 \rangle |^2 .
\ee
This expression can be Taylor expanded to second order in $\Delta t$
(assuming $\Delta t$ is small enough) taking into account that
\be
 \langle \Psi_0 | \Psi_{\Delta t} \rangle \approx
 1 - \frac{i\tau_Z}{\hbar}\ \langle \Psi_0 | \hat{H} | \Psi_0 \rangle
 + \frac{\tau_Z^2}{2\hbar^2}\ \langle \Psi_0 | \hat{H}^2 | \Psi_0 \rangle ,
\ee
where we have assumed the wave function is initially normalized.
This renders
%
%
\be
 \alpha_n = | \langle \Psi_0 | \Psi_{t_1} \rangle |^2 \approx
  1 - \frac{(\Delta t)^2}{\hbar^2}
    \left( \langle \Psi_0 | \hat{H}^2 | \Psi_0 \rangle
    - \langle \Psi_0 | \hat{H} | \Psi_0 \rangle^2 \right)
  = 1 - \frac{(\Delta t)^2}{\hbar^2}\ (\Delta \hat{H})^2 ,
\ee
which is valid for any $n$, since it does not depend explicitly on
the particular time at which the measurement is made.
Therefore, (\ref{photo}) can be expressed as
\be
 P(t) = \alpha_1^{n-1} |\langle \Psi_{t_{n-1}} | \Psi_t \rangle|^2 ,
\ee
which is formally equivalent to (\ref{survn}) after performing $n$
measurements, although it describes a completely different physics
\cite{tomas:2011}.




\end{document}